\documentclass[pra,aps,showpacs,superscriptaddress,twocolumn]{revtex4}
\usepackage{color} 
\usepackage{times}
\usepackage{graphicx}
\usepackage{epstopdf}
\usepackage[english]{babel}
\usepackage{mathrsfs}
\usepackage{textcomp}
\usepackage{amsthm}
\usepackage{amsfonts}
\usepackage{amssymb}
\usepackage{hyperref}
\usepackage{soul}
\usepackage{amssymb,amsmath,bbm,amsfonts,amsthm,subfigure,dsfont,eqnarray}
\usepackage{makecell}
\usepackage{setspace}

\newcommand{\abs}[1]{\left\vert#1\right\vert}

\newcommand{\ket}[1]{\left\vert#1\right\rangle}

\newcommand{\average}[1]{\left\langle #1 \right\rangle}

\begin{document}

\title{Local Quench, Majorana Zero Modes, and Disturbance Propagation in the Ising chain}

\author{G. Francica}
\affiliation{Dipartimento di Fisica, Universit\'a della Calabria,
87036 Arcavacata di Rende (CS), Italy} \affiliation{INFN - Gruppo
Collegato di Cosenza}

\author{T. J. G.  Apollaro}
\affiliation{NEST, Istituto Nanoscienze-CNR and Dipartimento di Fisica e Chimica, Universit$\grave{a}$  degli Studi di Palermo, via Archirafi 36, I-90123 Palermo, Italy}

\author{N. Lo Gullo}
\affiliation{Dipartimento di Fisica, Universit\`a degli Studi di
Milano, via Celoria 16, 20133, Milano, Italy}

\author{F. Plastina}
\affiliation{Dipartimento di Fisica, Universit\'a della Calabria,
87036 Arcavacata di Rende (CS), Italy} \affiliation{INFN - Gruppo
Collegato di Cosenza}

\date{\today}

\begin{abstract}
We study the generation and propagation of local perturbations in
a quantum many-body spin system. In particular, we study the Ising
model in transverse field in the presence of a local field defect
at one edge. This system possesses a rich phase diagram with
different regions characterized by the presence of one or two
Majorana zero modes. We show that their localized character {\it i})
enables a characterization of the Ising phase transition through a
local-only measurement performed on the edge spin, and {\it ii})
strongly affects the propagation of quasiparticles emitted after
the sudden removal of the defect, so that the dynamics of the
local magnetization show clear deviations from a ballistic
behavior in presence of the Majorana fermions.
\end{abstract}

\maketitle The impressive progress made in the last two decades in
the manipulation and detection of ultracold atomic gases,
\cite{bloch}, has had a decisive r\^ole in pushing towards a
better understanding of the dynamics of many body systems
following a sudden quench, i.e. an abrupt change of some control
parameter of the system. This has been the subject of many recent
studies, focusing in particular on global quenches \cite{review}.
Examples of long studied systems and processes which have been
realized with ultracold gases, range from the superfluid to Mott
insulator transition \cite{cazalilla} to the BCS to BEC
crossover\cite{Regal95PRL05,zwerger}. A special emphasis in this
respect deserve the studies of equilibration properties of
interacting many body systems \cite{gogolin}, including many
interesting results such as the transition from diffusive to
ballistic propagation  dynamics of boson in a one dimensional
lattice, \cite{Ronzheimer110PRL13}. The latter experiment is
connected to one of the paradigm in the theory of global quenches
in short ranged interacting many body systems, namely the
existence of a maximum finite speed for the propagation of
information within the system \cite{lrbound,Calabrese96PRL06}. On
the other hand, a general understanding is still lacking in the
case of local quenches, despite some interesting analysis for
local bond quenches \cite{locainterest}, for their connection to
orthogonality catastrophe \cite{OC}, for the study of local bound
states in interacting systems \cite{boundstates}, and for a recent
work by Smacchia and Silva~\cite{Silva109PRL12}, discussing the
propagation of magnetization after a local time dependent quench
in the quantum Ising model.

In the same spirit, and motivated by the increasing experimental
ability to perform single-site resolved addressing and detection
\cite{localb}, we study how a local quench affects both the static
and dynamical properties of an Ising chain of $N$ spins in a
transverse field $h$, which is homogenous everywhere but for a
local defect at the boundary.

For a homogeneous field and in the thermodynamic limit, the Ising
system would display a second order Quantum Phase Transition (QPT)
at a critical value of the external field $h=h_c$, separating a
paramagnetic phase with a non-degenerate ground state, from an
interaction dominated, ordered phase, with a two-fold degenerate
ground state. By means of a Jordan-Wigner transformation, the spin
system can be mapped onto an homogeneous Kitaev chain
\cite{Kitaev44PU01}, whose phase diagram has been studied in
detail~\cite{degottardi2013} and is known to display a
topologically non-trivial phase, and a trivial one. The
non-trivial phase is characterized by the presence of boundary
Majorana fermions, which have been extensively studied both
theoretically~\cite{Liu2006,Fu2008,Jiang2011} and
experimentally~\cite{Akhmerov2008,Das2012,Mourik2012}, also with
the aim of exploiting them to implement topological quantum
computation\cite{nayak}. Global quench induced dynamics have been
addressed for this system to study the transport of localized
excitations between the two edges of the
chain~\cite{degottardi2011,Vasseur2014}, a necessary step to
implement topological gates.

The magnetic field defect we consider gives a twofold effect: 1)
the topological Majorana mode in the non-trivial phase gets
distorted; 2) a localized eigenmode appears, which itself becomes
a zero mode within a one-dimensional subregion of the otherwise
trivial topological phase. As a consequence, the phase diagram in
presence of the defect becomes much richer, due to the presence of
either one or two of these localized modes. Finally, when the
defect is quenched off, quasiparticles are ``emitted'' from a
finite region around it and propagate throughout the system
showing a clear signature of the presence of the Majorana mode.

\section{Model and phase diagram} 
\begin{figure}[ht]
        \begin{center}
        \includegraphics[width=\linewidth]{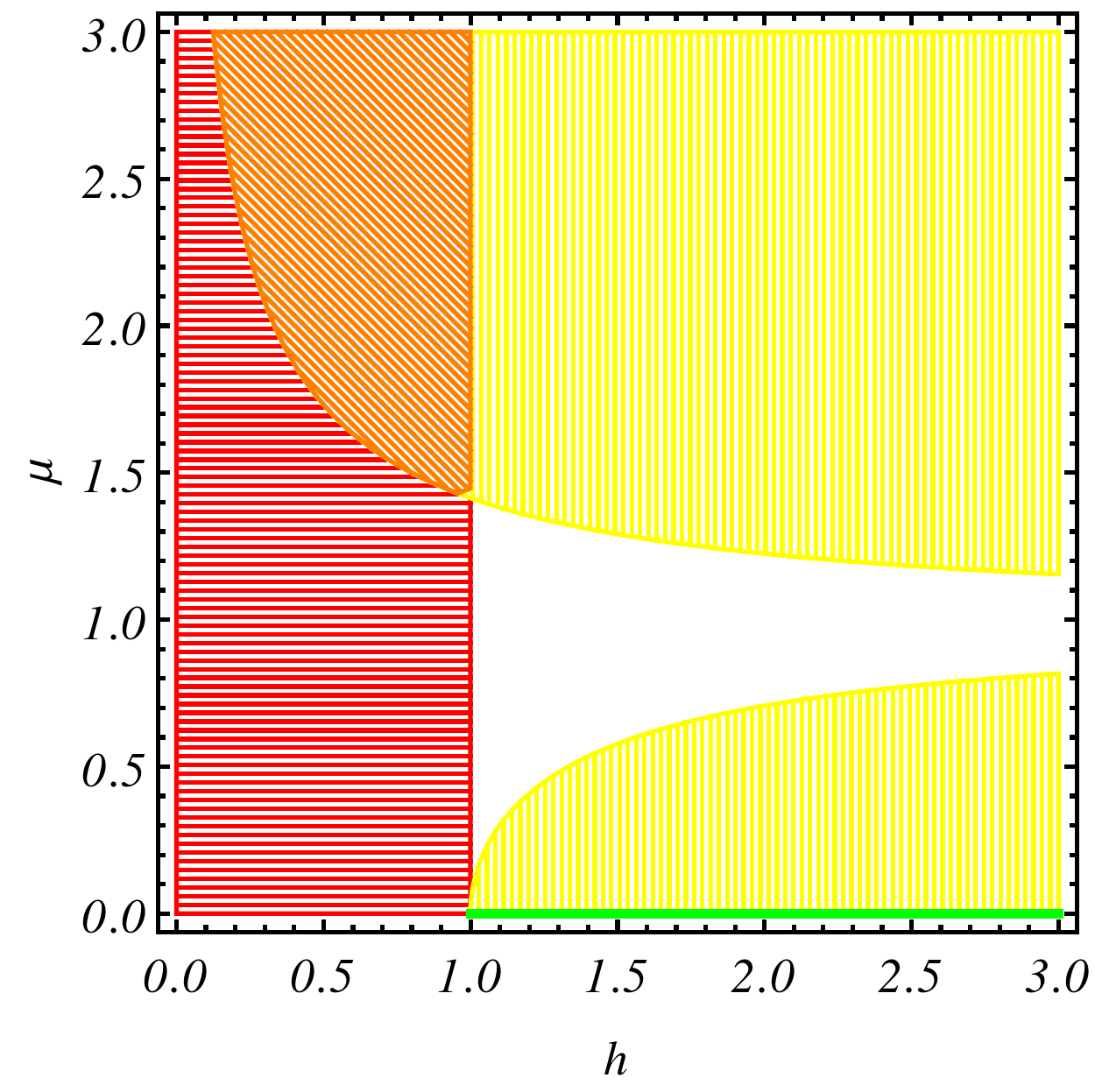}
        \caption{(Color online):
        Phase diagram showing the presence of the localized modes $\Lambda_{n}$ ($n{=}1,2$) in the $h,\mu$ plane.
        The red region (horizontal lines), ${\cal R}_1$, features the presence of the mode with energy
        $\Lambda_{1}$. The yellow region (vertical lines) ${\cal R}_2$, instead, features the presence of mode $2$, either below or above the band in the two
        subregions with $\mu$ smaller or larger than $\mu=1$, respectively.
        Both of the localized modes are present in the orange region (oblique lines), corresponding to the intersection of ${\cal R}_1$ and ${\cal R}_2$.
        On the solid green line ($\mu=0$ with $h\geq 1$) we have
        $\Lambda_{2}=0$ and mode $2$ becomes a zero mode. Finally, only delocalized modes are present in
        the white region.}
        \label{Figdiscr}
        \end{center}
\end{figure}

We consider the transverse field
Ising model with open boundary conditions in the presence of a
local dip in the magnetic field (a {\it defect}), described by the
Hamiltonian:
\begin{equation}
H_\mu = - J \Bigl \{ h \sum_{n=2}^N \hat\sigma^z_n -
\sum_{n=1}^{N-1} \hat\sigma^x_{n}\hat\sigma^x_{n+1} + \mu h\;\hat
\sigma^z_1 \Bigr \}, \label{E.HIsing}
\end{equation}
where $\mu$ is used to parameterize the field defect, and we
scaled the magnetic field so that $h_c{=}1$. From now on, we take
the exchange constant as our energy unit, $J=1$. Despite the
breaking of both translation invariance and reflection symmetry,
$H_\mu$ can be diagonalized in terms of fermion operators
$\eta_k$, $\eta_k^{\dag}$, (the diagonalization, based on
\cite{vecchi,Yueh2015,BanchiVaia13}, is discussed in the Appendix)
 \begin{equation} \label{hdiag} H_\mu= \sum_{\kappa}
\Lambda_{\kappa} \eta_{\kappa}^{\dag} \eta_{\kappa} + \chi_1
\Lambda_{1} \eta_1^{\dag} \eta_1 + \chi_2 \Lambda_{2}
\eta_2^{\dag} \eta_2 \, ,\end{equation} where $\kappa$ runs over a
quasi-continuous band of delocalized modes, while two further {\it
discrete} modes can appear depending on the value of the magnetic
field $h$ and of the defect parameter $\mu$. Let ${\cal R}_{n}$ be
the region in parameter space where mode $n=1,2$ exists; then,
${\cal R}_1$ is the ferromagnetic region $h\leq 1$, while ${\cal
R}_2=\{(h,\mu): (\forall h
\wedge|\mu|{>}\sqrt{1{+}1/h})\;{\vee}\;(h>1\wedge|\mu| {<}
\sqrt{1{-}1/h})\}$, see Fig. (\ref{Figdiscr}) and the Appendix. In
Eq. (\ref{hdiag}), $\chi_1 = \Theta (1-h)$ and $\chi_2$ are the
characteristic functions of these two regions, so that the
corresponding fermion mode $n=1$ ($n=2$) is absent if $h, \mu$ are
taken outside ${\cal R}_1$ (${\cal R}_2$). These two modes, have
frequency
\begin{equation}\Lambda_{1} {=} \frac{2
\mu(1{-}h^2)h^{N}}{\sqrt{\abs{1{+}(\mu^2 {-}1) h^2 }}} \, , \quad
\Lambda_{2} = 2 \abs{\mu} \sqrt{\frac{1{+}(\mu^2-1)h^2}{(\mu^2-1)}}
\, .
\end{equation}
Mode $1$ originates from fermion pairing, as found by Kitaev
\cite{Kitaev44PU01} for a homogeneous system. In our case, it is
distorted by the presence of the defect (i.e. if $\mu\ne 1$), both
in its energy and in its spatial structure. It becomes a zero mode
in the thermodynamic limit, remaining spatially localized on the
boundaries of the chain for any value of $\mu$. Mode $2$
originates from the defect, it is discrete and localized too; its
energy can lie either below or above the band (lower or upper of
the yellow subregions in Fig. \ref{Figdiscr}, respectively). If
$\mu=0$, $\Lambda_2$ becomes zero even at finite size; as a
result, the real fermion operators $\eta_2^{(a)}$ and
$\eta_2^{(b)}$ defined as $\eta_2 = \eta_2^{(a)}+ i \eta_2^{(b)}$
decouple from the Hamiltonian, so that mode $2$ becomes a Majorana
zero mode, localized around the defect (see Table~(IA) in the
Appendix).

\begin{figure}[ht]
        \begin{center}
        \includegraphics[width=\linewidth]{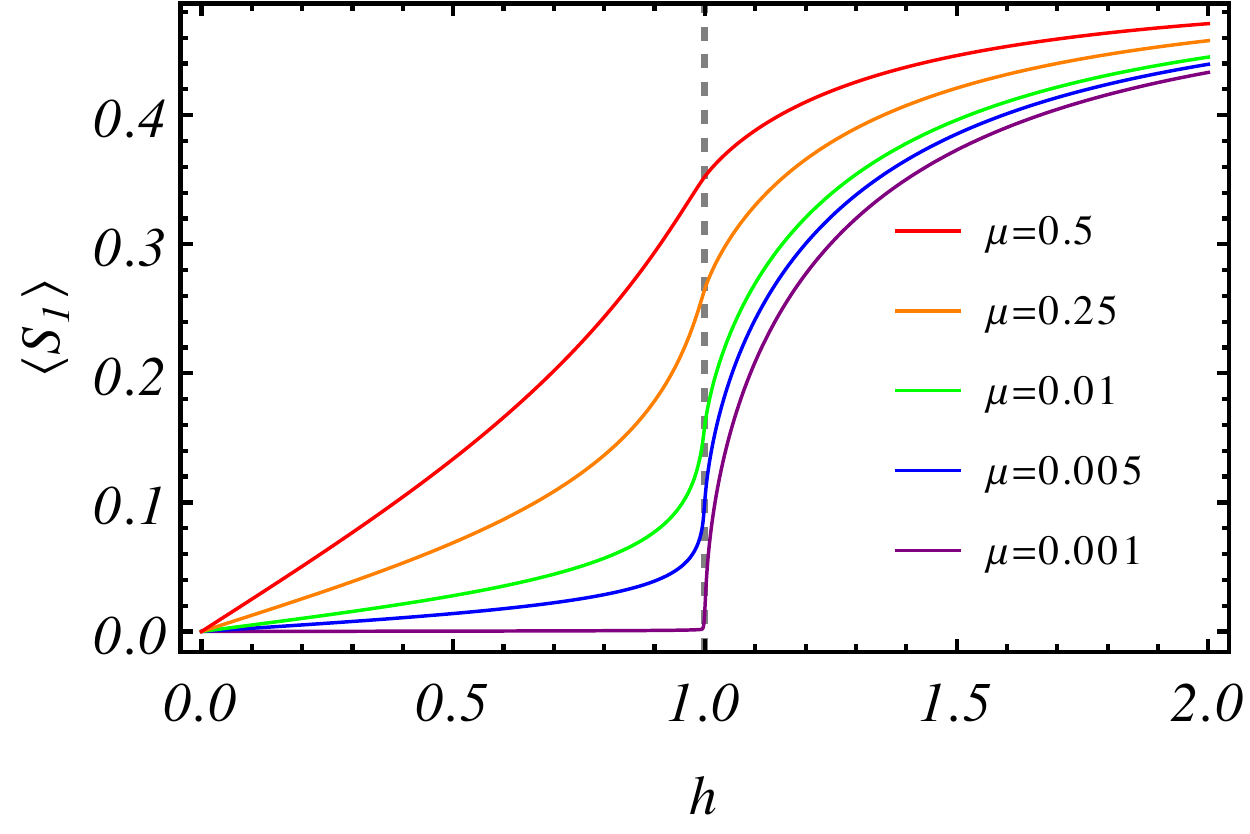}
\caption{(Color online): Transverse magnetization of the defect
spin as a function of the magnetic field for various defect\rq{}s
strength $\mu$ (with $\mu{=}1$ corresponding to a homogeneous
system), for $N=1000$. }
        \label{F.nhomo}
        \end{center}
\end{figure}

The presence of such localized structures affects all of the
static properties of the system. In particular, we will focus on
the local transverse magnetization that, in the Jordan-Wigner
language, is related to fermion occupation. As shown in
Fig.~\ref{F.nhomo}, the defect-localized mode maintains a
non-vanishing magnetization on the first site for every finite
value of $\mu$. In particular, $\left \langle S_1^z \right
\rangle$  grows linearly with the external magnetic field with a
slope proportional to $\mu$ far from criticality. For $\mu
\rightarrow 0$, however, we obtain a singular behavior,
$\displaystyle\lim_{\mu{\to}0^\pm}\!\!\average{
S^z_1}{=}\mp\frac{\sqrt{h^2{-}1}}{2h} \Theta(h-1)$, whose
step-like nature originates from a discontinuity in the spatial
structure of mode $2$ in the paramagnetic region, see also the
Appendix. The fact that the magnetisation is zero (for vanishing
$\mu$) in the ferromagnetic region can be qualitatively justified
by observing that the spin-spin interaction locally dominates in
this case, preventing the impurity spin to acquire a finite
magnetisation in the $z$-direction. On the contrary, in the
paramagnetic regime, the one-body Hamiltonian term dominates in
Eq.~\ref{E.HIsing}, resulting in a building-up of
$\average{S^z_1}$ even for vanishingly small $\mu$.

Because of these features, the impurity spin and its magnetization
behave as a local probe, able to detect the bulk properties of the
spin chain: for $\mu=0$, $\average{S^z_1}$ is zero in the ordered
phase, while it is different from zero in the disordered one. This
is quite peculiar as, in general terms, for a second-order QPT as
the one we are facing here, critical properties are exhibited in
the bulk,
and no {\textit{local}} measurement of the transverse
magnetization or of its susceptibility close to the boundary is
able to pinpoint the QPT. On the other hand, we have just shown
that the local magnetization on the edge impurity site is able to
capture and signal the QPT.

\section{Propagation of quasi-particles} We now turn to the
study of the dynamics following the sudden removal of the defect
on the first site. The aim is to discuss how a local perturbation
propagates in the system and, in particular, how the dynamics is
affected by the  Majorana zero mode. To this end, we assume the
system to be initially prepared in the ground state $\ket{GS}_0$
of $\hat H_{0}$ (with $\mu=0$). At $t=0$ the defect is suddenly
removed ($\mu=1$ for $t\ge 0$), so that the system's subsequent
evolution is generated by the homogeneous Ising Hamiltonian $\hat
H_1$, whose ground state we denote $\ket{GS}_1$.

\begin{figure}[ht]
        \begin{center}
\includegraphics[width=\linewidth]{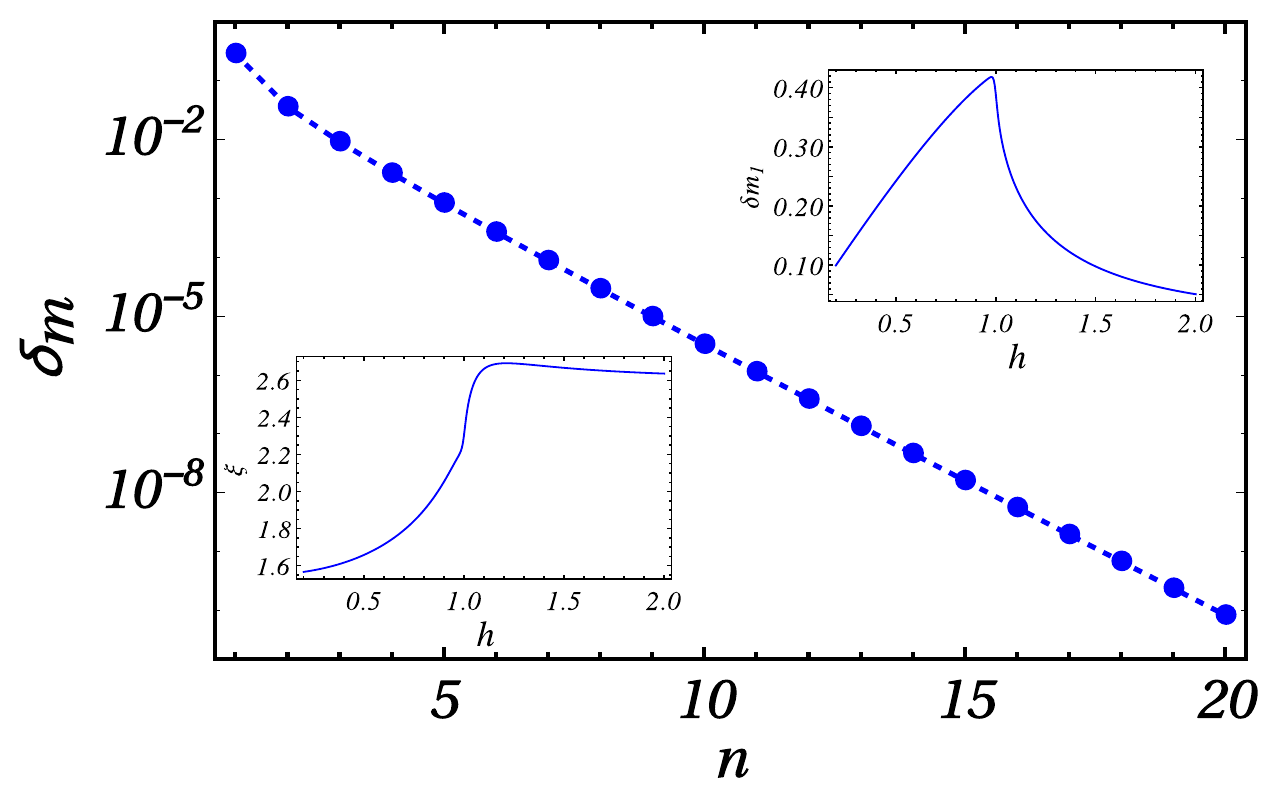}
\caption{(Color online): Magnetization change $\delta m$ between
the ground states of the final and initial Hamiltonians for
$N=200$ spins and $h=0.6$. Insets: (top) magnetization change at
the first site, $\delta m_1$, as a function of $h$; (bottom)
localization length $\xi$. }
        \label{Figlength}
        \end{center}
\end{figure}

The spatial structure of the initial state $\ket{GS}_0$ differs
from that of $\ket{GS}_1$ near the first site only. To
characterize the local differences between these two states we
employ the magnetization contrast $\delta m_i = \langle \hat S_i^z
\rangle_{GS_1} - \langle  \hat S_i^z \rangle_{GS_0}$, which
quantifies the defect-induced perturbation of the ground state
with respect to the homogeneous field Hamiltonian $\hat H_1$ at a
given $h$. We expect the two ground states to look very similar
far from the defect site and to differ significantly only around
it. Indeed, $\delta m_i$ decays exponentially with the distance
from the defect, $\delta m_i = \delta m_1 \exp(-(i-1)/\xi)$, with
a short localization length $\xi$, see Fig.\ref{Figlength}.
Notice, in particular, that the perturbation is always localized
within the first three sites regardless of the value of $h$. On
the other hand, $\delta m_1$ increases with $h$ for $0<h<1$ while
it goes to zero for $h>1$ away from the critical point (indeed, in
the paramagnetic phase, the magnetization tends to saturate with
increasing $h$, both with and without the defect).

Once the defect is removed, the local magnetization peak travels
through the chain, starting near site $i=1$ at $t=0$. In fact, we
can think of the region of size $\xi$ around the first site as a
source of quasi-particles that carry magnetization and
correlations \cite{Calabrese96PRL06,Silva109PRL12}.

Two different scenarios occur, depending on the value of the
transverse field $h$. For $h>1$, after the quench the system only
supports delocalized fermion eigen-modes (white region in
Fig.\ref{Figdiscr}). These will be shown to give rise to a purely
ballistic propagation of the magnetization peak. In the ordered
phase $0<h<1$, on the other hand, $H_1$ enjoys the localized mode
with energy $\Lambda_1$, residing on the edge of the system, and
substantially overlapping with the initial localized state
$\ket{GS}_0$.
A pinning of the excitation near the first site occurs in this
case, due to the interplay of the otherwise ballistic propagation
with the localized nature of the Majorana mode, and giving rise to
temporal oscillations of the local magnetization.

In order to characterize the propagation of the magnetic
perturbation along the chain, we consider the mean square
magnetization center and its velocity, defined as
\begin{eqnarray}
\label{eq:rt}
  R^2(t) &=& \sum_{i=1}^{N} \delta m_i(t)(i-1)^2\\
  \label{eq:vt}
  v(t) &=& \frac{d}{dt} \sqrt{\abs{ R^2(t)-R^2(0)}}.
\end{eqnarray}
where $\delta m_i(t) = \langle \hat S_i^z \rangle_{GS_1} - \langle
\hat S_i^z(t) \rangle_{GS_0}$ is the time-dependent version of the
magnetization contrast introduced above. Analogous variables have
been adopted and experimentally measured in Ref.
\cite{Ronzheimer110PRL13}.

\begin{figure}
 \begin{tabular}{cc}
  \includegraphics[width=4cm]{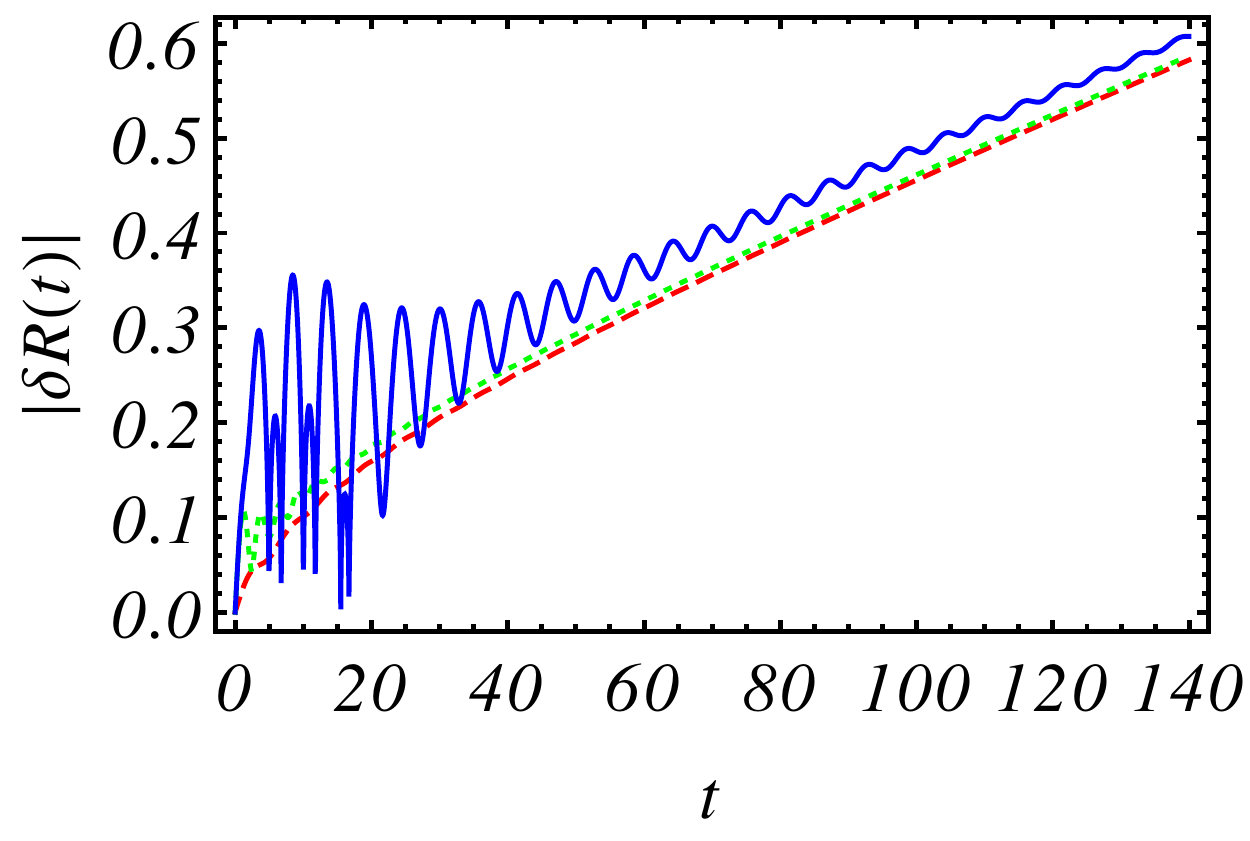}&\includegraphics[width=4cm]{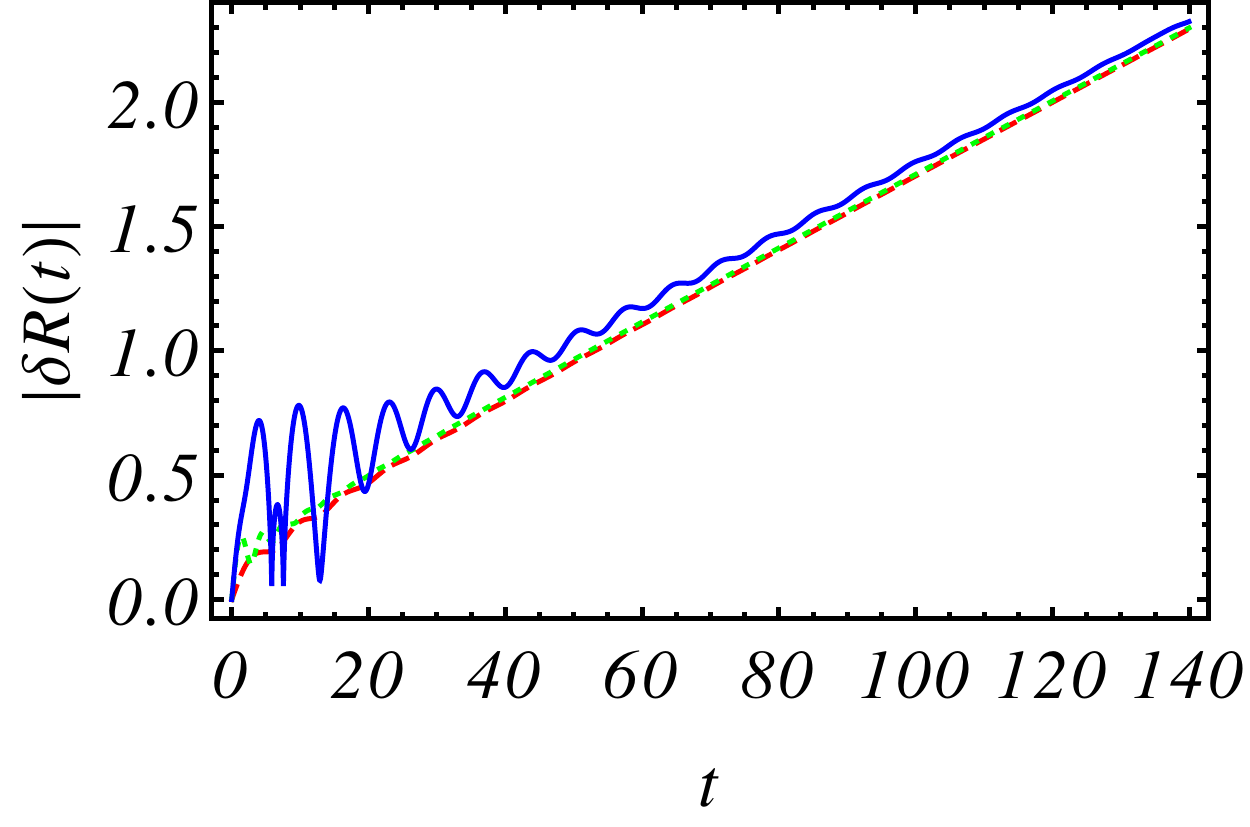}\\
  \includegraphics[width=4cm]{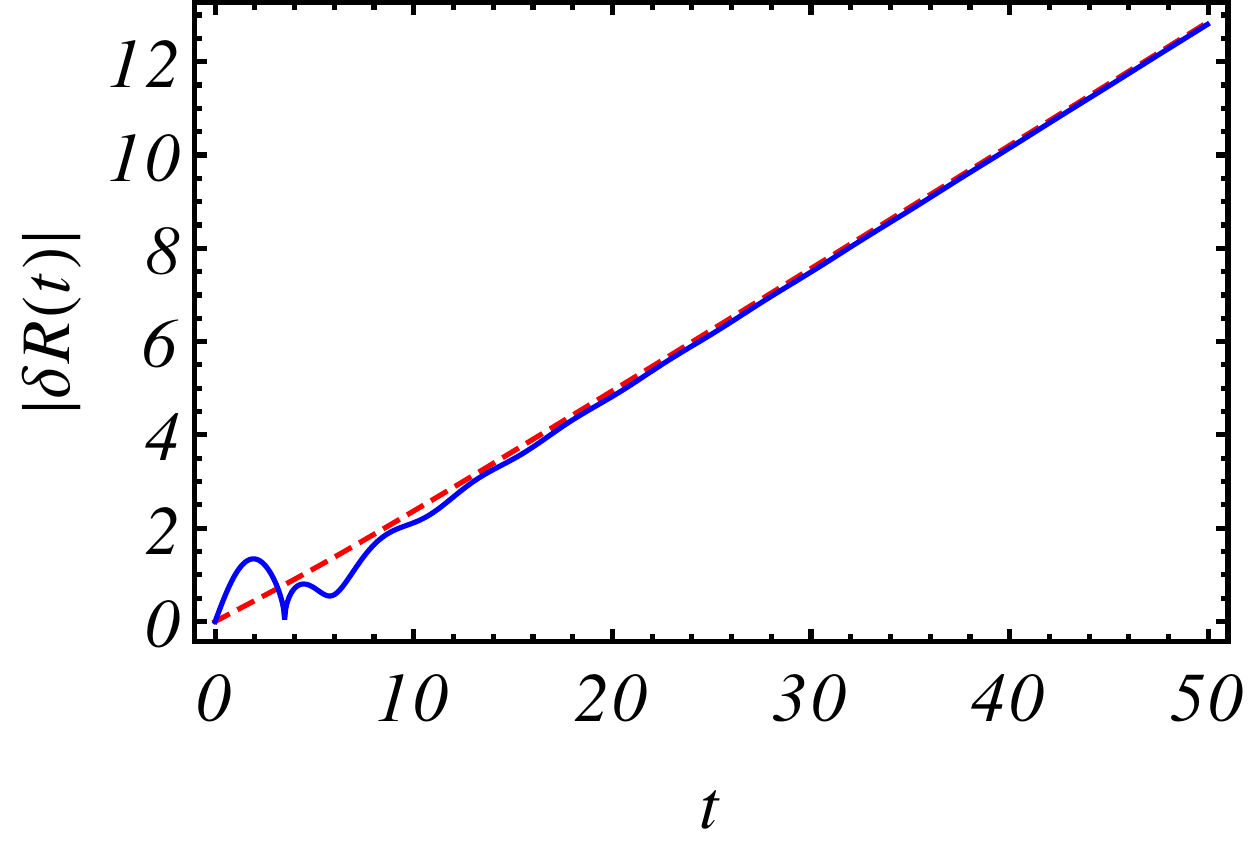}&\includegraphics[width=4cm]{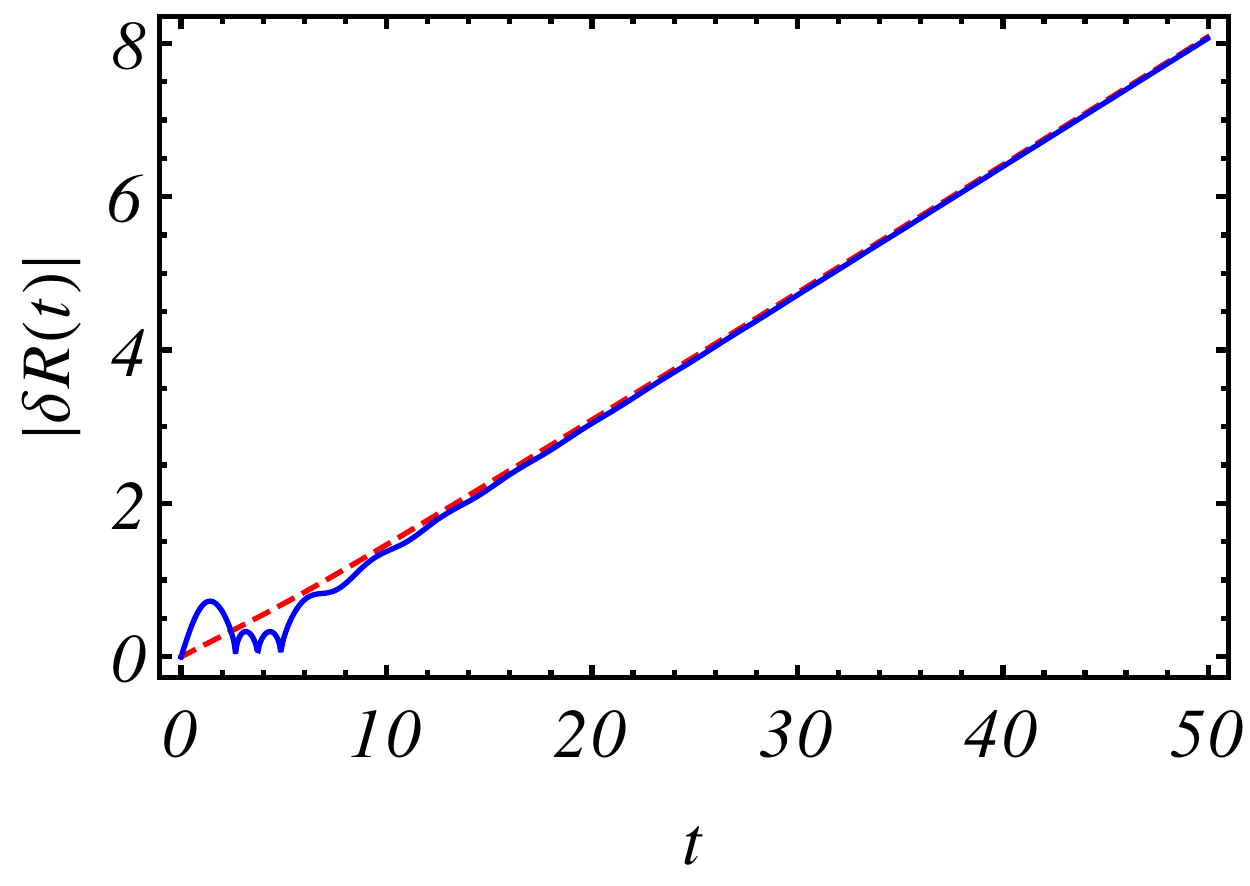}\\
 \end{tabular}
\caption{(Color online): Detachment of the magnetization center
from its initial position, $\delta R(t)=\sqrt{R^2(t)-R^2(0)}$, for
(top left to bottom right) $h=0.4,0.6,1.2,1.6$, with $N=100$.
Solid blue lines are drawn by using the full expression for
$R(t)$, whereas dashed red lines contain the rotating $(A)$
contributions only. The dashed green lines  are obtained from the
blue ones by artificially excluding the contribution from the
Majorana zero mode, in order to better highlight its role.}
\label{fig:dR}
\end{figure}

\begin{figure}
 \begin{tabular}{cc}
  \includegraphics[width=4cm]{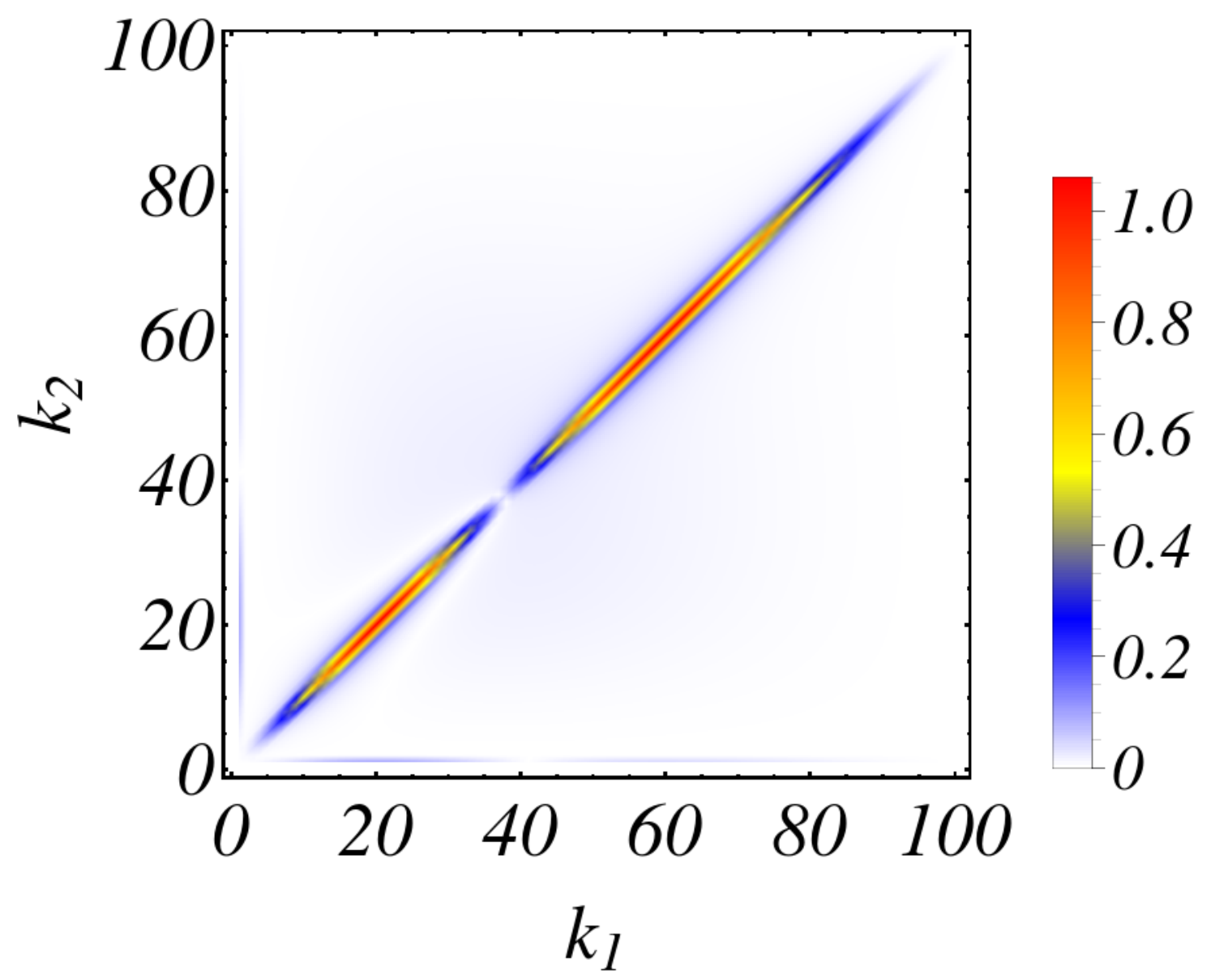}&\includegraphics[width=4cm]{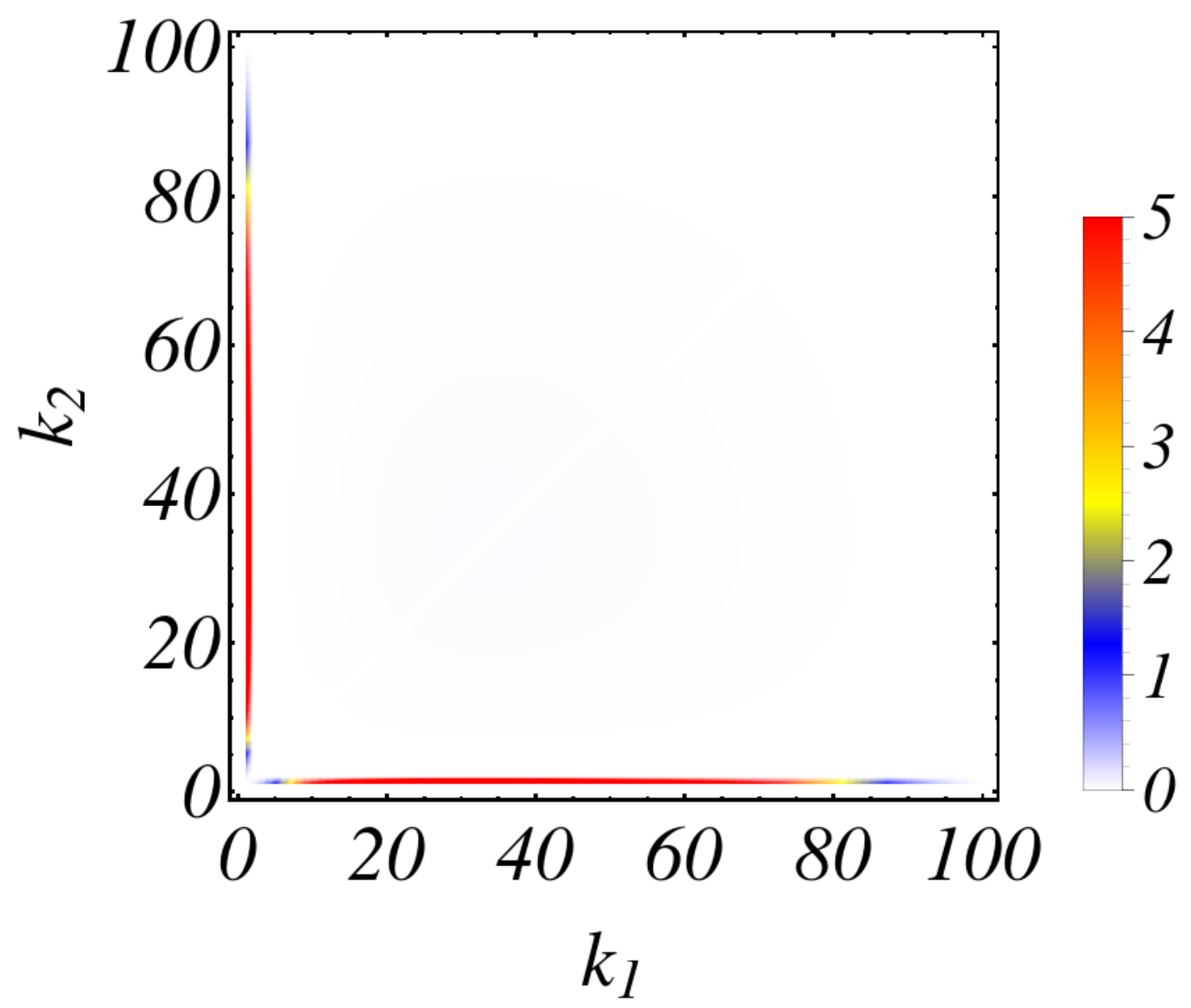}\\
  \includegraphics[width=4cm]{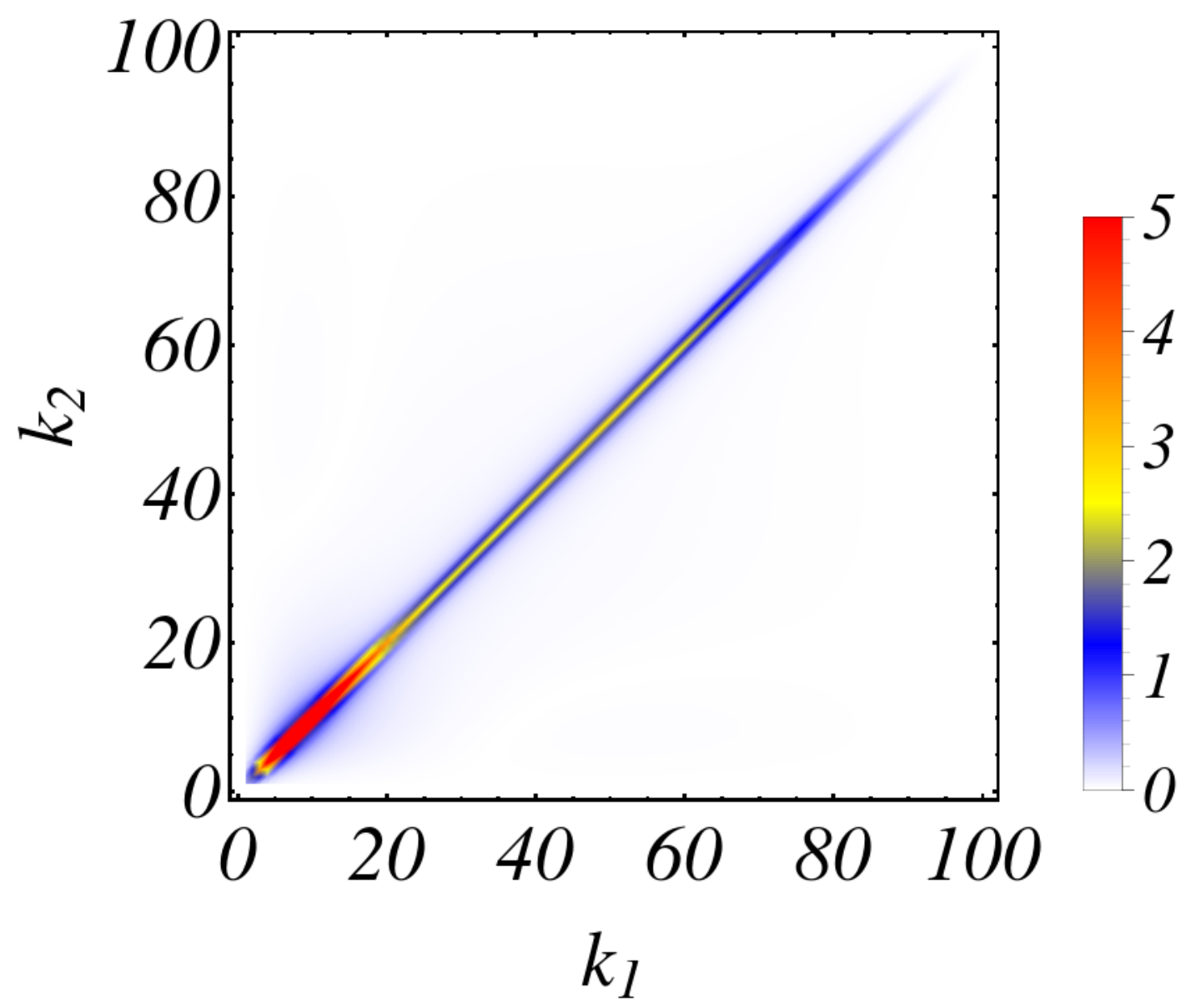}&\includegraphics[width=4cm]{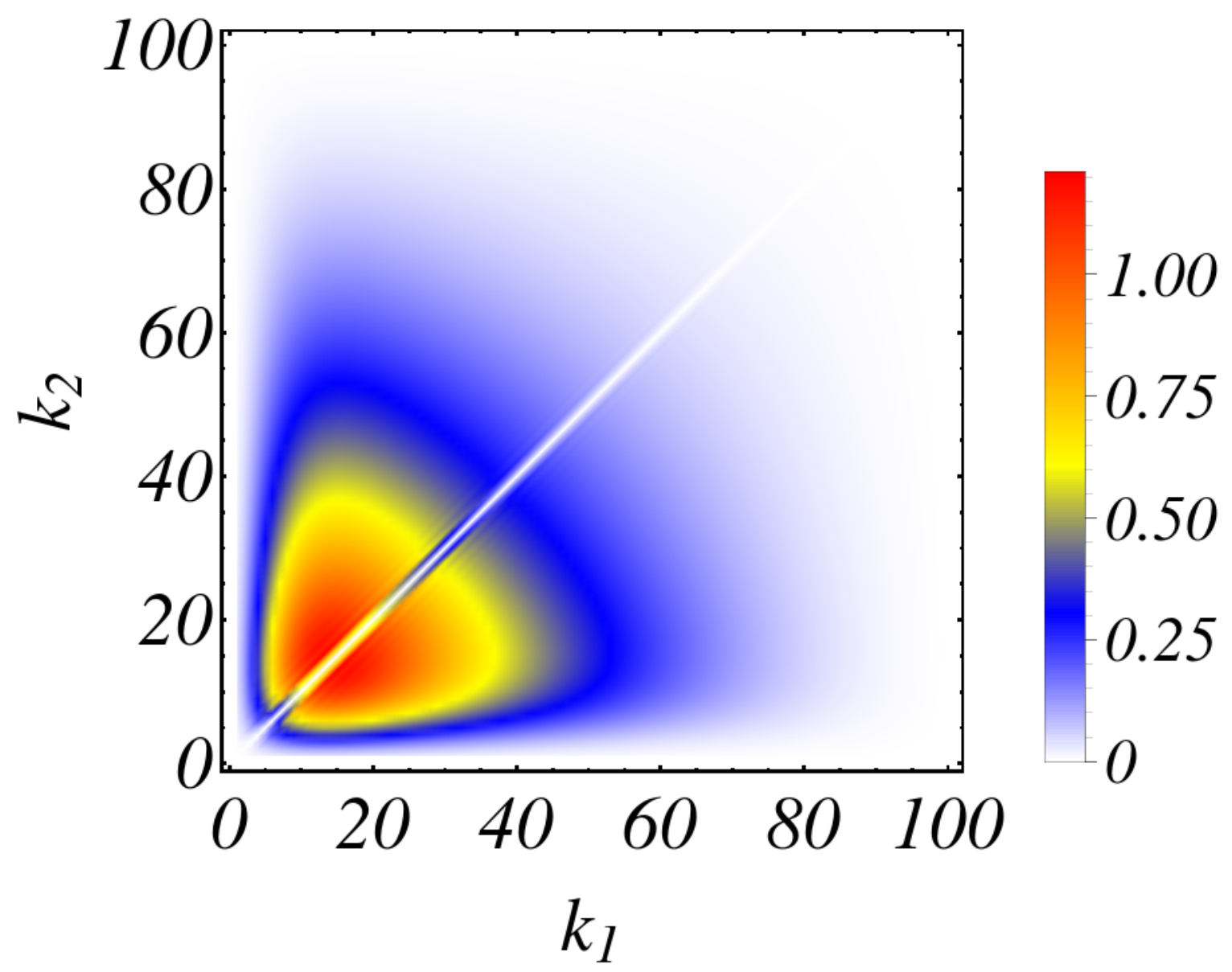}\\
 \end{tabular}
\caption{(Color online): Elements of the matrices $A$ (left
panels) and $B$ (right ones) for  $h=0.4$ (top) and $h=1.4$
(bottom). The plots are made for $N=100$.} \label{fig:AandB}
\end{figure}

\begin{figure}[ht]
\begin{center}
\begin{tabular}{cc}
  \includegraphics[width=4cm]{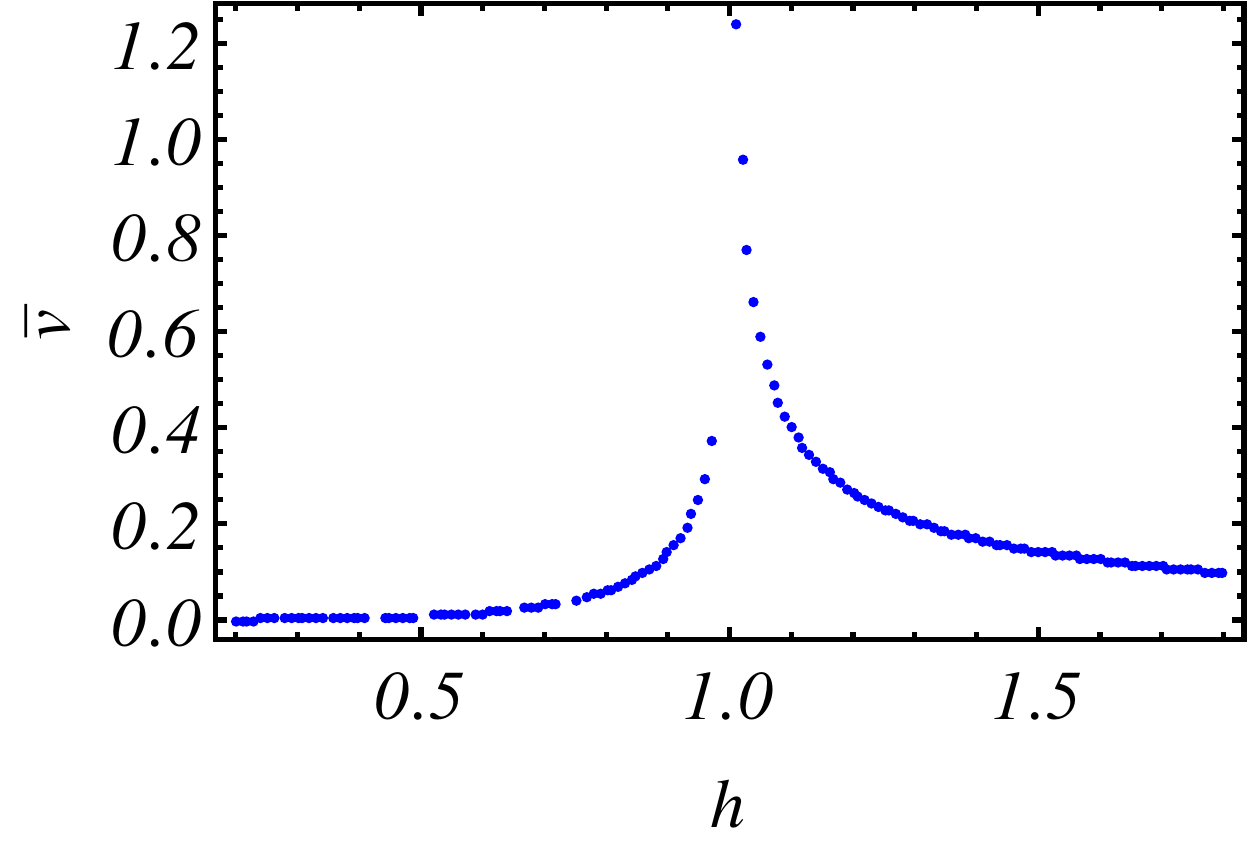}&\includegraphics[width=4cm]{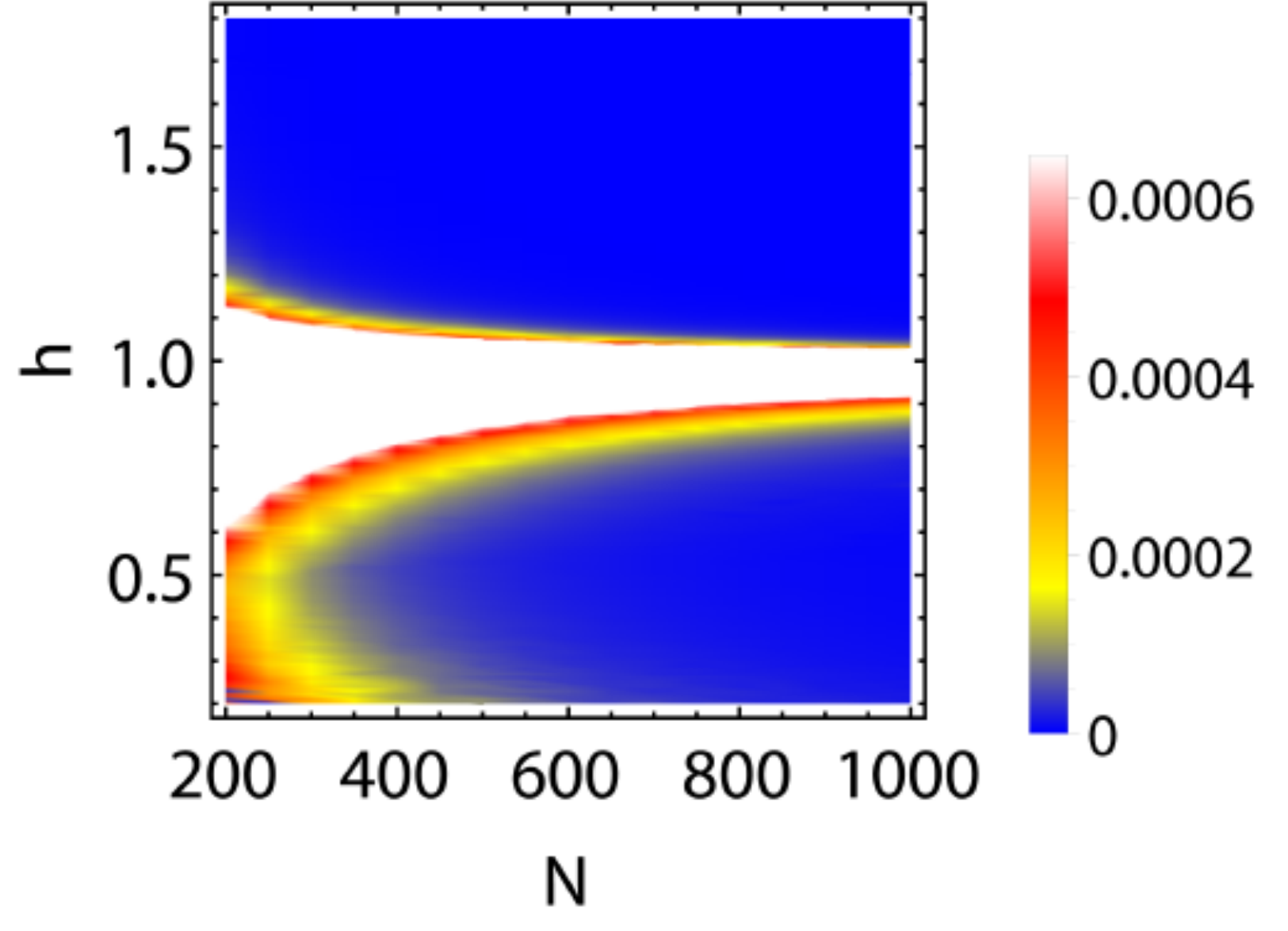}\\
  \end{tabular}\caption{(Color online): Left: Average asymptotic speed $\bar v$,
taken in the long time limit (but before the occurrence of finite
size revivals) as a mean over residual long time oscillations, for
$N=800$. Right: Scaling with the system size $N$ and with the
external field $h$ of the amplitude of the long time oscillations
displayed by the propagation speed $v(t)$ around the average $\bar
v$. In the white region the limiting speed is not well defined
because of the persistence of its oscillations.}
        \label{velo}
        \end{center}
\end{figure}

Using the diagonal form of $H_1$, one can show that
\begin{eqnarray}
  R^2(t) &=& \sum\limits_{k_1,k_2} A_{k_1\,k_2} \cos\left((\Lambda_{k_1}-\Lambda_{k_2})t\right)\nonumber\\
   &+& \sum\limits_{k_1,k_2}B_{k_1\,k_2} \cos\left((\Lambda_{k_1}+\Lambda_{k_2})t\right)\nonumber
\end{eqnarray}
where the summations are performed over the eigenmodes of the
final Hamiltonian $H_1$, including both the delocalized fermion
modes $\kappa$ forming a band in the thermodynamic limit, and, if
$h<1$, the (Majorana) edge mode $n=1$. The explicit form of the
matrices $A$ and $B$ are given in the Appendix; what is important
here is that they give two different types of contribution to the
propagating magnetization center: a rotating term, $(A)$, and a
counter-rotating one, $(B)$. The former fully determines the
asymptotic behavior of $R(t)$, as the $B$-contribution becomes
negligible at long times due to their fast oscillations. This is
clearly seen in Fig.\ref{fig:dR}, where we show $\delta
R(t)=\sqrt{R^2(t)-R^2(0)}$ for different values of the magnetic
field $h$. In the plots, the solid blue curves giving $\delta
R(t)$ are compared to the behaviors obtained by artificially
keeping the rotating terms only in the rhs of Eq.(\ref{eq:rt}).
This is done to better emphasize that $B$-terms only contribute to
the transient oscillations, after which the propagation is
ballistic and completely accounted for by the rotating terms.
Furthermore, by analyzing the matrix $A$ as displayed in Fig.
(\ref{fig:AandB}), we see that the main contribution comes from
the entries close to the diagonal, so that the long-time speed is
$$\bar{v} \approx \sqrt{\sum_{\kappa} \frac{A_{\kappa \, \kappa +
1}}{2} \left ( \Lambda_{\kappa}-\Lambda_{\kappa + 1} \right )^2}
\, .$$ This gives a very good approximation for the average
propagation velocity in the disordered region, while it fails near
the critical point, where $\delta R(t)$ and $v(t)$ keep
oscillating even at long times, see Fig. (\ref{velo}), and in the
ordered region, because of the presence of the Majorana mode,
coming into play via the counter rotating terms.

In fact, the $B$ contributions are worth analyzing in some detail.
They are basically irrelevant for $h>1$, while their presence
induces strong transient oscillations in the ferromagnetic phase,
whose amplitude increases as $h\rightarrow 1^{-}$. This is
essentially due to the presence of the Majorana mode. Indeed, Fig.
(\ref{fig:AandB}) clearly shows that in the ordered phase ($h<1$),
the only contribution of the $B$-type comes from the coupling
between the $n=1$ mode and the delocalized ones. Furthermore, in
Fig. (\ref{fig:dR}), for $h<1$ (top panels) we can see that, if
all of the terms involving the Majorana zero mode were excluded
from the sum in Eq. (\ref{eq:rt}), then the oscillations would
disappear (as seen by comparing the dotted green curves with the
solid blue ones). Therefore, it is the presence of the Majorana
mode in the final Hamiltonian that gives rise to the large and
persistent fluctuations in $\delta R(t)$, which can be understood
as a result of the localized mode tieing the magnetization peak at
short times.

\section{Conclusions} To summarize, we studied the effect of
a local magnetic field defect on both the static and dynamic
properties of the Ising model in transverse field $h$. The
excitation spectrum of this system is made of a continuous band of
delocalized states and, depending on the interplay between the
external field $h$ and the defect parameter $\mu$, of up to two
localized modes.  We have shown that for $0<h<1$ the defect
modifies the wave function of the (Kitaev) fermion paring induced
Majorana  mode, and that it can give rise to a new localized zero
mode in the disordered region; furthermore, we demonstrated that,
in the limit of a vanishing magnetic field on the edge, the Ising
critical point can be detected by means of local measurement on
the impurity. We have also studied the propagation of local
magnetic excitations occurring after the defect is quenched off
and showed that the propagation is ballistic but for some
oscillations induced by the the Majorana mode for $0<h<1$, whose
effect persist at long times near the critical point.

{\it Acknowledgments -}  We are grateful to T. Alecce and L.
Dell'Anna for helpful discussions, and acknowledge financial
support by the EU Collaborative project QuProCS (Grant Agreement
641277).  T.J.G.A. acknowledges funding  under the EU Collaborative Project TherMiQ (Grant No. 618074)

\begin{appendix}
\section{}
\subsection{Diagonalization of the Ising model with an edge
defect}\label{Ss.Isingdia} The Ising model with a magnetic field
inhomogeneity at one edge is given by
 \begin{equation}\label{E.IsingH}
 \hat H = -\mu h\; \hat\sigma_1^z-h \sum_{n=2}^{N} \hat\sigma^z_n+\sum_{n=1}^{N-1} \hat\sigma^x_n \hat\sigma^x_{n+1}~,
 \end{equation}
where $\hat\sigma^{\alpha}_n$ ($\alpha{=}x,y,z$) are the usual
Pauli spin operator on site $n$, and $h>0$. Diagonalization of
Eq.~\ref{E.IsingH} is achieved first by introducing the non-local
Jordan-Wigner (J-W) transformation \cite{vecchi}
\begin{eqnarray}
 \hat \sigma_n^z &=& 2\hat c_n^\dag \hat c_n-1,\\
 \hat \sigma_n^- &=& e^{ i \pi \sum_{j=1}^{n-1}\hat c_j^\dag \hat c_j } \hat c_n,\\
 \hat \sigma_n^+ &=& e^{ -i \pi \sum_{j=1}^{n-1}\hat c_j^\dag \hat c_j } \hat c_n^\dag,
\label{E.JW}
\end{eqnarray}
where $\hat\sigma^{\pm}_j{=}(\hat\sigma_j^x\pm i
\hat\sigma_j^y)/2$ are the raising and lowering spin operator.
This transformation fermionizes Eq.~\ref{E.IsingH} into
 \begin{equation}
 \hat H = \sum_{i\, j} \left(\hat c^\dag_i A_{i\,j}\hat c_j + \frac{1}{2}\left(\hat c^\dag_i B_{i\,j} \hat c^\dag_j + h.c. \right) \right)
\label{E.IsingJW}
 \end{equation}
where $A$ and $B$ are tridiagonal symmetric and anti-symmetric
matrices respectively, whose elements are given by $A_{i\,j}{=} 2
h \left((\mu+1)\,
\delta_{i1}{-}1\right)\delta_{ij}{-}\left(\delta_{i\,j{+}1}{+}\delta_{i{+}1\,j}\right)$
and $B_{i\,j}
=\left(\delta_{i\,j{+}1}{-}\delta_{i{+}1\,j}\right)$. Since the
Hamiltonian in Eq.~\ref{E.IsingJW} is bilinear in the creation and
annihilation operators it can be diagonalized by means of a
Bogoliubov transformation:
\begin{equation}
\hat c_i=\sum_k u_{ik}\hat \eta_k+v_{ik} \hat \eta^\dag_k~
\label{E.Bogo}
\end{equation}
with the conditions
$\sum\limits_{k}u_{ik}u_{jk}+v_{ik}v_{jk}=\delta_{ij}$ and
$\sum\limits_{k}u_{ik}v_{jk}+v_{ik}u_{jk}=0$ to ensure that the
transformation is canonical and preserves the anti-commutation
relations. From the equations of motion for the operators $\hat
c_i$ (or equivalently for $\hat c_i^{\dag}$), the Bogoliubov
transformation in Eq.\ref{E.Bogo} and imposing the time dependence
$\eta_k(t)=\eta_k e^{-\imath\Lambda_k t}$ for the normal modes we
obtain the following equations for the element of the
transformation matrices $u$ and $v$:

\begin{eqnarray}
\sum\limits_{j}A_{ij}u_{jk}+B_{ij}v_{jk}=\Lambda_k u_{ik}\\
\sum\limits_{j}B_{ij}u_{jk}+A_{ij}v_{jk}=-\Lambda_k v_{ik}.
\end{eqnarray}
The Hamiltonian rewritten in terms of the normal modes reads $\hat
H=\sum\limits_{k}\Lambda_k\hat\eta_k^{\dag}\hat\eta_k-N h +
(\mu+1)h-\frac{1}{2}\sum\limits_{k}\Lambda_k$.

In view of the discussion about the Majorana modes we introduce
the new matrices $\phi=u^T+v^T$ and $\psi=u^T-v^T$ whose column
vectors satisfy the equations:

\begin{eqnarray}
\begin{cases}
\left(A+B\right)\vec{\phi}_k{=}\Lambda_k \vec{\psi}_k\\
\left(A-B\right)\vec{\psi}_k{=}\Lambda_k \vec{\phi}_k\\
\end{cases}~.
\end{eqnarray}

{\textit{In the absence of impurity}} the above equations can be
decoupled, yielding respectively:
\begin{equation}
M_{1}\vec{\psi}_k{=} \Lambda_k^2 \vec{\psi}_k~~~\text{or}~~~
M^{'}_1\vec{\phi}_k {=} \Lambda_k^2 \vec{\phi}_k~,
\label{E.Lambda}
 \end{equation}
where $M_1{=}(A{+}B)(A{-}B)$ and $M_1^{'}{=}(A{-}B)(A{+}B)$ turns
out to be the mirror-inverted matrix $M_1^{'}{=}R^{\dagger}M_1 R$,
with $R$ denoting the reflection operator
$R_{i\,j}{=}\delta_{i\,2N{+}1{-}j}$. Because of the open boundary
conditions, $M_1$ ($M_1^{'}$) are symmetric tridiagonal matrices
uniform along the diagonals but for the first (last) element on
the main one. The position of this non-uniformity makes it
possible to determine analytical solutions for the spectrum of
$M$~\cite{Yueh2015,BanchiVaia13}.

The presence of the impurity $\mu$ breaks the mirror-inversion
symmetry and therefore the diagonalization procedures for
$M_{\mu}$ or  $M'_{\mu}$, although numerically possible even for
large $N$, are different. In fact, $(A{-}B)(A{+}B)$ turns out to
be a real, tridiagonal matrix with the upper corner
({\textit{i.e.}}, the matrix elements $\{(1,1),(1,2),(2,1)\}$)
depending on $\mu$, and this, to the best of our knowledge, does
not admit an analytical solution for arbitrary values of $\mu$.
The matrix $M_{\mu}{=}(A{+}B)(A{-}B)$, instead, has only the first
element on the main diagonal depending on $\mu$ (beside a
$\mu$-independent non-uniformity on the last element of the same
diagonal). As a consequence, the latter is prone to analytical
diagonalization for arbitrary values of the impurity strength by
means of the technique outlined in Ref.~\cite{Yueh2015}.

The matrix $M_{\mu}{=}(A+B)(A-B)$ is positive, tridiagonal, and
symmetric, and reads
 \begin{equation}
 M =  \left(
   \begin{array}{ccccc}
     b-\alpha & a & 0 & \cdots & 0 \\
     a & b & a & \cdots & 0 \\
     0 & a & b & \cdots & \cdots \\
     \cdots & \cdots & \cdots & \cdots & \cdots \\
     0 & \cdots & a & b & a \\
     0 & \cdots & 0 & a & b-\beta \\
   \end{array}
 \right)~,
\end{equation}
with $b=4+4h^2$, $a=-4h$, $\alpha = 4h^2 (1-\mu^2)$, $\beta= 4$.
Lengthy, but straightforward, calculations lead to the following
equation
 \begin{equation}
 a(\alpha {+} \beta) \sin(N \theta_k){+}a^2 \sin \left[ (N{+}1)\theta_k \right] {+} \alpha \beta \sin \left[ (N{-}1)\theta_k \right] {=} 0
\label{eig val eq}
 \end{equation}
where $\theta_k{\in} {\mathds C}$ is related to the eigenvalues
$\Lambda_k$  of Eqs.~\ref{E.Lambda} by $\cos\theta_k {=}
\frac{\Lambda_k^2{-} b}{2a}$, with the index $k$ labelling the
energy levels. Finally we solve Eq.~\ref{eig val eq} in the limit
$N{\gg}1$. Having obtained the allowed $\theta_k$, the problem is
solved and the matrices $\psi$ and $\phi$ are obtained via
Eqs.~\ref{E.Lambda}.

Depending on the values of the Hamiltonian parameters $\{h,\mu\}$,
there can be up to two complex $\theta_k$, which give rise to the
out-of-band energy levels reported in the main text. The phase
diagram in the $\{h,\mu\}$-plane is made up of the following
regions: ${\cal R}_1{=}\{(h,\mu): 0 {<} h {<} 1\}$ and ${\cal
R}_2=\{(h,\mu): (\forall h
\wedge|\mu|{>}\sqrt{1{+}1/h})\;{\vee}\;(h>1\wedge|\mu| {<}
\sqrt{1{-}1/h})\}$
A graphical representation of these regions is given in the main
text. Results for the eigenvalues, and for the coefficients $\psi$
and $\phi$ are reported in Table~\ref{T.Sol}, for the various
regions.

\begin{widetext}

\begin{table}[htdp]\label{T.Sol}
\caption{Expressions for the eigenvalues and the $\{\psi,\phi\}$
matrix elements in the $(h,\mu)$-plane.}
\begin{center}
\begin{tabular}{|c|c|c|c|}
\hline
$(h,\mu)$ & $\Lambda$ & $\psi$ & $\phi$ \\
\hline \thead{$\forall(h,\mu)$} & \thead{$\Lambda_\kappa{=} 2
\sqrt{ 1 {+} h^2 {-} 2 h \cos\theta_\kappa }$ }

& \thead{$  \psi_n(\theta_\kappa){=} \sqrt{\frac{2}{N}}
\frac{\sin(n\theta_\kappa) {+} (\mu^2-1)h
\sin((n{-}1)\theta_\kappa)}{\sqrt{1+(\mu^2-1)^2 h^2 {+} 2h (\mu^2
{-}1)\cos\theta_\kappa}}$ } & \thead{ $
\phi_n(\theta_\kappa){=}\frac{2h}{\Lambda_\kappa}
\psi_n(\theta_\kappa)
{-}\frac{2(\mu+1)h\delta_{n1}}{\Lambda_\kappa}
\psi_1(\theta_\kappa){-}\frac{2(1{-}\delta_{n1})}{\Lambda_\kappa}
\psi_{n{-}1}(\theta_\kappa)$}
\\
\hline $\thead{{\cal R}_1}$ & $\thead{ \Lambda_{1} {=} \frac{2
|\mu|(1{-}h^2)h^{N}}{\sqrt{\abs{1{+}(\mu^2 {-}1) h^2 }}} }$ &
$\thead{ \psi^{(1)}_n {=} \sqrt{1{-}h^2}\left( h^{N{-}n} {-}
\frac{\mu^2 h^{N{+}n}}{1{+}(\mu^2-1)h^2} \right) }$ &

$\thead{\phi^{(1)}_n {=}
\frac{\sqrt{1{-}h^2}\sqrt{\abs{1{+}(\mu^2-1)h^2}}}{|\mu|
h(1{+}(\mu^2-1)h^2)} \left( 1 {-} (\mu+1)\left(
\delta_{n1}{+}(1{-}\delta_{n1})\left(1-\mu\right) \right)
\right)h^n}$
\\
\hline \thead{${\cal R}_2$} & {\thead{$\Lambda_{2} = 2 |\mu|
\sqrt{\frac{1{+}(\mu^2-1)h^2}{(\mu^2-1)}}$}} &
\thead{$\psi^{(2)}_n{=} \frac{({-}1)^n
h^{-n}\sqrt{(\mu^2-1)^2h^{2}{-}1}}{{\left(\mu^2-1 \right)}^{n}}$}
& {\thead{$\phi^{(2)}_n {=}\frac{2h}{\Lambda^{(2)}}  \psi^{(2)}_n
{-}\frac{2(\mu+1)h \delta_{n1}}{\Lambda^{(2)}}  \psi^{(2)}_1  {-}
2\frac{1{-}\delta_{n1}}{\Lambda^{(2)}} \psi^{(2)}_{n{-}1}$}}
\\
\hline
\end{tabular}
\end{center}
\end{table}

\end{widetext}

The first line of Table~\ref{T.Sol}, where
\begin{equation}
\theta_k {=} \frac{k\pi}{N}{+}\frac{1}{N} \arctan\left(
\frac{h\mu^2\sin\left(\frac{k \pi}{N}\right)}{((\mu^2 {-} 1 )h-1)
+ h(2-\mu^2) \cos\left(\frac{k \pi}{N}\right)} \right),
\label{E.Theta}
\end{equation}
with $k\in \mathds{Z}$, refers to spatially delocalized energy
modes having energy in the interval $E_b{=}
\left[2\abs{1{-}h},2\abs{1{+}h}\right]$. For finite systems, there
are $N,N{-}1$, or $N{-}2$ modes depending on location in the
$(h,\mu)$-plane. For infinite systems, on the other hand, they
build up an energy band with a gap equal to $2\abs{1{-}h}$ closing
at the QPT point $h_c{=}1$. The second line shows the zero-energy
mode (in the $N{\rightarrow}\infty$ limit) and the corresponding
Majorana localized eigenstate which is extensively discussed in
the main text. Finally, the last line reports an energy level
appearing above (or below) the band for $(h,\mu){\in} {\cal R}_2$.
The corresponding eigenstate turns out to be localized only around
the impurity and has a finite energy but on the line $\mu{=}0$
where it vanishes $\forall h$. The occurrence of
$\Lambda^{(2)}{=}0$ implies a discontinuity for $\phi_1^{(2)}$ and
a non-differentiable point for $\phi_n^{(2)}$ with $n{>}1$.
Indeed, $\phi^{(2)}_1{=}- \frac{2 \mu h }{ \Lambda^{(2)}}
\sqrt{1{-}((\mu^2-1)h)^{{-}2}}$ and  $\phi^{(2)}_{n}{=} \frac{ 2 h
\mu^2}{ \Lambda^{(2)}}{\left( (\mu^2-1)h \right)}^{{-}n}
\sqrt{{\left( (\mu^2-1)h \right)}^{2}{-}1}$.

By introducing two (real) Majorana operators on each site: $\hat
a_{n} =\hat c_n +\hat c^\dag_n$ and $\hat b_{n} = i(\hat c^\dag_n
-\hat c_n)$ together with the Bogoliubov transformations in
Eq.\ref{E.Bogo} we have
\begin{eqnarray}
  \hat a_{n} &=& \sum_\kappa \phi_{n}(\theta_\kappa)\hat \alpha_\kappa + \sum_{i} \phi^{(i)}_n \hat\alpha_i\\
  \hat b_{n} &=& \sum_\kappa \psi_{n}(\theta_\kappa)\hat \beta_\kappa + \sum_{i} \psi^{(i)}_n\hat \beta_i~
\end{eqnarray}
where we defined the Majorana modes $\hat \alpha_k=\hat
\eta_k^\dag+\hat\eta_k$ and $\hat \beta_k= i (\hat\eta_k^\dag-\hat
\eta_k)$ and the functions $\phi$ and $\psi$ are given in
Table~(\ref{T.Sol}).

The Hamiltonian thus reads:
\begin{equation}
  \hat H_\mu = i \mu h\; \hat a_{1}\hat b_{1}+i h \sum_{n=2}^{N} \hat a_{n}\hat b_{n} - i\sum_{n=1}^{N-1} \hat a_{n+1}\hat b_{n}.
\label{eq:majorham}
\end{equation}
In this form it is easy to see that for $\mu=0$ the Majorana
operators $\hat a_1$ and $\hat b_1$ both commute with the
Hamiltonian. 

\subsection{Transverse magnetization}\label{Ss.TM} The
magnetization is given by $\average{\hat
S^z_n}{=}\frac{\average{\hat
\sigma^z_n}}{2}{=}\frac{1}{2}{-}\average{\hat c_n^{\dagger}\hat
c_n}$. By inverting Eq.~\ref{E.Bogo} and exploiting the property
$\eta_k\ket{GS}{=}0$, $\forall k$, we are left with $
\average{\hat S^z_n}{=}\frac{1}{2}-\sum_k v_{kn}^2$. In the
thermodynamic limit $N{\rightarrow}\infty$, we obtain the exact
result
\begin{eqnarray}\label{E.magn}
&\average{\hat S^z_n}{=}\frac{1}{2}{-}\left[\int_0^{\pi}d\theta\,
v^2_n(\theta){+}\Theta\left(h{-}1\right)\Theta\left(x^+\right)
\Theta\left(y^+\right) \left(v_n^{(2)}\right)^2\right.\\
&\left.+\Theta\left(1{-}h\right)\left(
\left(\Theta\left(y^-\right){+}\Theta\left(x^-\right)\right)
\left(v_n^{(2)}\right)^2{+}
\left(v_n^{(1)}\right)^2\right)\right]\nonumber~,
\end{eqnarray}
where $v_n^{(i)}$ is the Bogoliubov coefficient relative to the
discrete mode $i=1,2$, while
$x^{\pm}{=}\mu{\pm}\sqrt{h(h{\mp}1)}$,
$y^{\pm}{=}{\pm}\sqrt{h(h{\mp}1)}{-}\mu$, and $\Theta(x)$ is the
Heaviside step function.

Let us focus on the case $\mu{\rightarrow}0$. Eq.~\ref{E.magn} can
be evaluated analytically as the contribution of the continuous
modes in the integrand, given by
\begin{align}\label{E.sz_cont}
v^2_n(\theta)&=\frac{1}{2 \pi f^2}
\left[{-}h\, \widetilde{\sin}\left(\left(n{-}2\right)\theta\right)\right. \\
&\left.{+}\left(1{+}h^2{-}h\, f
\right)\widetilde{\sin}\left(\left(n{-}1\right)\theta\right){+}\left(f{-}h\right)\widetilde{\sin}\left(n\theta\right)\right]\nonumber~,
\end{align}
where $f{\equiv}f(h,\theta){=}\sqrt{1{+}h^2{-}2h\cos\theta}$ and
the modified trigonometric function reads
${\widetilde{\sin}}(x){=}\Theta(x)\sin(x)$, results in rational
functions of $h$ times complete elliptic integrals of the first
and second kind for $n{\geq}2$. For the sake of brevity, we do not
report the result for arbitrary $n$ of Eq.~\ref{E.magn}, but,
rather, focus on the case $n{=}1$, which reads
\begin{align}\label{E.mag1}
&\displaystyle\lim_{\mu{\to}0^{\pm}}\!\!\average{\hat S^z_1}{=}\frac{1}{2}{-}\frac{1}{8h^2}\left(1+Sg\left(h{-}1\right){+}h^2\left(1+Sg\left(1{-}h\right) \right)\right)\nonumber\\
&{-}\Theta\left(h{-}1\right)
 \frac{2h\left(h{\mp}\sqrt{h^2{-}1} \right){-}1}{4 h^2}{-}\frac{\Theta\left(1{-}h\right)}{4}
 \nonumber \\ &{=}\mp\frac{\sqrt{h^2{-}1}}{2h} \Theta(h-1)~,
\end{align}
where $Sg(x)$ is the sign function.

\subsection{Coupling matrices}\label{sec:ABs} In the first section
of this Appendix, we have shown how to diagonalize an Hamiltonian
having the form in Eq. \ref{E.IsingJW}. When performing a sudden
quench both the initial and the final Hamiltonian have this form
and can thus be diagonalized following the above procedure. To
this end, we need two Bogoliubov transformations as in
Eq.\ref{E.Bogo}:

\begin{eqnarray}
\hat c_i=\sum_i u_{ik}\hat \eta_k+v_{ik} \hat \eta^\dag_k~\\
\hat c_i=\sum_i w_{ik}\hat \xi_k+z_{ik} \hat \xi^\dag_k~
\end{eqnarray}
where again we require
$\sum\limits_{k}u_{ik}u_{jk}+v_{ik}v_{jk}=\delta_{ij}$ and
$\sum\limits_{k}w_{ik}w_{jk}+z_{ik}z_{jk}=\delta_{ij}$. Using
these relations and the transformations above we can write:

\begin{equation}
\left(
\begin{array}{c}
\hat\xi_k\\
\hat\xi_k^\dag
\end{array}
\right)=\sum_k\left(
\begin{array}{cc}
U_{kq}& V_{kq}\\
V_{kq}& U_{kq}
\end{array}
\right)\left(
\begin{array}{c}
\hat\eta_q\\
\hat\eta_q^\dag
\end{array}\right)
\end{equation}
where
\begin{eqnarray}
U_{kq}=\sum_i w_{ik} u_{iq}+z_{ik} v_{iq}\\
V_{kq}=\sum_i z_{ik}u_{iq}+w_{ik}v_{iq}.
\end{eqnarray}
Thanks to the above expression it is now easy to recover the
expression in the main text for the mean local number of fermions
at site $i$:

\begin{eqnarray}
\langle \hat c_i^\dag(t)\hat c_i(t)\rangle&=&\sum\limits_{k_1,k_2}A_{k_1k_2}^{(i)}\cos\left(\left(\Lambda_{k_1}-\Lambda_{k_2}\right)t\right)\\
&+&\sum\limits_{k_1,k_2}B_{k_1k_2}^{(i)}\cos\left(\left(\Lambda_{k_1}+\Lambda_{k_2}\right)t\right)
\end{eqnarray}
where
\begin{eqnarray}
A_{k_1k_2}^{(i)}&=&(w_{ik_1}w_{ik_2}-z_{ik_1}z_{ik_2})\sum\limits_{l}V_{k_1l}V_{k_2l}\\
B_{k_1k_2}^{(i)}&=&2w_{ik_1}z_{ik_2}\sum\limits_{l}U_{k_1l}V_{k_2l}.
\end{eqnarray}

The matrices $A$ and $B$ in the definition of $R^2(t)$ in the main
text are given by

\begin{eqnarray}
A_{k_1k_2}&=&\sum\limits_i A_{k_1k_2}^{(i)} (i-1)^2\\
B_{k_1k_2}&=&\sum\limits_i B_{k_1k_2}^{(i)} (i-1)^2.
\end{eqnarray}
\end{appendix}

\end{document}